\documentclass[twocolumn,showpacs,preprintnumbers,amsmath,amssymb]{revtex4}
\usepackage{amsfonts}
\usepackage{amsmath}
\usepackage{amssymb}
\usepackage{subfigure}
\usepackage{graphicx}
\begin{document}

\title[Magneto-resistance of BSCCO]{Free flux flow magneto-resistance of single crystal BSCCO }
\author{A.Pallinger$^{1}$}
\author{G.Kriza$^{1}$}
\author{B.Sas$^{1}$}
\author{I.Pethes$^{1}$}
\author{K.Vad$^{2}$}
\author{F.I.B.Williams$^{1,3}$}
\affiliation{$^{1}$SZFKI Research Institute for Solid State Physics and Optics, PO Box 49,
H-1525 Budapest, Hungary}
\affiliation{$^{2}$Institute of Nuclear Research, PO Box 51, H-4001 Debrecen, Hungary}
\affiliation{$^{3}$CEA-SaclaySerivce de Physique de l'Etat Condens\'{e}, Comissariat \`{a}
l'Energie Atomique, Saclay, F-91191 Gif-sur-Yvette, France}
\pacs{74.72.Hs, 74.25.Fy, 74.25.Qt, 74.25.Sv}
\date{\today}

\begin{abstract}
Measurement of the free flux flow resistance in monocrystalline highly
anisotropic BSCCO reveals a low field magneto-resistive effect nearly $100$
times that expected from a naive application of the Bardeen-Stephen rule
$R/R_{n}=B/B_{c2}$. Furthermore in the vortex solid phase it saturates to a
constant value at about $H_{c2}/100$ as if the vortices were moving $100$
times faster than expected. An attempt to account for this behavior by a
combination of sheared planar vortex flow obeying the BS rule and the
quasi-particle $c$ axis conductivity between the Josephson coupled
superconducting planes is only partially successful. In particular it is
unable to account for the saturation feature which occurs in both optimally
and underdoped samples in the low temperature vortex solid phase.

\end{abstract}

\maketitle

\section{Introduction}

High T$_c$ cuprate superconductors differ from their low T$_c$ counterparts in their
extreme type II character with a Ginzburg-Landau parameter $\kappa
\thicksim200$ corresponding to extremely short coherence length $\xi\thicksim1$ nm
and high upper critical fields $B_{c2}\thicksim100$ T. The small Abrikosov vortex core size of order $\xi$ results in a series of
bound states with energies situated in the gap whose mean energy spacing is of
order $<\delta\epsilon>\thicksim\Delta^{2}/E_{F}\thicksim\hbar^{2}/m\xi^{2}$
where $\Delta$ is the superconductor gap energy (modulus of the order
parameter), $E_{F}$ the Fermi energy and $m$ the effective mass. The core
structure can affect the mobility for translational motion of the vortices by
modifying the spectral properties of the electron momentum transfer.

Dissipation in type II superconductors in magnetic field $H>H_{c1}$ in slowly
varying electromagnetic fields is dominated by the dynamics of Abrikosov
vortices \cite{kopnin}. Material details of the superconductor enter mainly
via the \textquotedblleft friction coefficient\textquotedblright\ that gives
the vortex velocity in terms of the driving Magnus-Lorentz force. The friction
is controlled by the spectral density and relaxation properties of the
low-lying \textquotedblleft core states\textquotedblright, i.e., quasiparticle
states localized to the vortex core. In dirty superconductors, where the mean
free path $\ell\ll\xi$, the friction coefficient is related to the normal state
resistivity by the Bardeen-Stephen (BS) rule \cite{bardeen1965}:
\begin{equation}
\rho_{f}=\alpha\rho_{n}B/B_{c2}\ ,
\label{eq:bs}
\end{equation}
where $\rho_{f}$ is the flux flow resistivity arising from vortex motion in
the absence of pinning, $\rho_{n}$ is the normal state resistivity, $B$ is the
magnetic field, $B_{c2}$ the upper critical field and $\alpha=1$. A more
careful examination of the core states allowed the extension of Eq. (\ref{eq:bs}) with
$\alpha\approx1$ to the moderately clean limit $\ell>\xi$ as well
\cite{kopnin}. The BS law has been experimentally confirmed for a broad range
of conventional ($s$-wave) superconductors \cite{parks}.

In unconventional superconductors with gap nodes, BCS theory suggests a high
density of core states at the Fermi energy. Nevertheless calculation
\cite{kopnin1997} of the flux-flow resistivity revealed that as long as one
stays in the moderately clean limit, Eq. (\ref{eq:bs}) remains valid with a possible
enhancement of $\alpha$ but still of order $1$. Recent measurements
\cite{alpha} on several anisotropic non-high-$T_{c}$ superconductors that most
likely exhibit gap nodes confirm Eq. (\ref{eq:bs}) with a moderately enhanced
$1.6<\alpha<4.7.$

The problem of low-temperature vortex dynamics in BSCCO (Bi$_{2}$Sr$_{2}%
$CaCu$_{2}$O$_{8+\delta})$ is particularly interesting in this respect in the
light of the recent results on the structure of the vortices in this material.
Scanning tunneling microscope (STM) spectroscopy \cite{maggio-aprile} revealed
that the zero-energy peak in the density of core states is missing. This,
together with the results of inelastic neutron scattering \cite{ins} and NMR
\cite{nmr} experiments, led to the suggestion that some concurrent, non
superconducting, order exists at the vortex cores when the superconducting
order parameter is suppressed \cite{kv}. The idea received direct support from
STM spectroscopy\cite{hoffman} where a periodic modulation of the local
density of electronic states around the vortex cores was observed. Since the
decay length of this modulation is much longer than the superconducting
coherence length, not only the structure of the core, but also the structure
of the flow field for vortex transport should be different from conventional
Abrikosov vortices with probable consequences for the velocity-force relation.

The experimental situation in the high-$T_{c}$ superconductors, and in
particular in BSCCO, is not clear. Low-frequency transport measurements
\cite{bs-bscco} in the vortex liquid phase close to the critical temperature
$T_{c}$ are in reasonable agreement with the BS law. Microwave and millimeter
wave impedance measurements \cite{microwave} at low temperatures, on the other
hand, indicate a large and nearly $B$-independent dissipation in a broad field range.

To clarify the situation and to see if these very special aspects of the
vortex structure influence the dynamics, we have performed a systematic
investigation of the flux-flow resistance in BSCCO single crystals by
measuring the $a$-$b$ plane voltage-current ($V$-$I$) characteristics up to
currents well above the threshold current for dissipation in $c$-directed
magnetic field. The free flux-flow resistance as measured on an $ab$ face can
be approximately described as $\propto B^{1/2}$ for low fields followed by
saturation (becomes field independent) above about 1 T in the low-temperature
vortex solid phase. As the saturated value corresponds quite  well with the
extrapolation of the normal resistance to low temperatures, it might be
naively interpreted as reflecting $\alpha\sim100$ in Eq. (\ref{eq:bs}), at least at 1 T,
indicating that vortices move at up to two orders of magnitude faster than
predicted by BS law in low fields. Although it is known that vortex velocities
much larger than the BS value may result from a nonlinear instability at high
vortex velocities \cite{larkin}, not only does its onset in BSCCO
films\cite{bs-bscco} occur at current densities about two orders of magnitude higher
than investigated in this study, the characteristic signature
\cite{bs-bscco,kunchur} of a nonlinear runaway in the $V$-$I$ curve is absent in our case.

\section{Experiment}

Measurements were made on nine single crystals from three different batches of
Bi$_{2}$Sr$_{2}$CaCu$_{2}$O$_{8+\delta}$ fabricated by a melt cooling
technique \cite{sample}. The typical dimensions of the crystals were
$1\times0.5\times0.003$ mm$^{3}$ with the shortest dimension corresponding to
the poorly conducting $c$ axis. Most of the crystals have been close to
optimal doping with a resistance-determined critical temperature $T_{c}%
\approx89$~K and transition width about 2~K in zero field. The diamagnetism in
a $10~$Oe field set in in a uniformly progressive way below $T_{c}$ to near
$100\%$ at low temperature. Two underdoped crystals exhibited a resistive
$T_{c}\approx51$~K and $T_{c}\approx53$~K with width 6-8 K and slow initial
diamagnetism onset to about $30~$K before proceeding to about $40\%$ flux
exclusion at low temperature.

The resistance measurements were made in the usual four point configuuration
on an $ab$ face of the crystals. The contacts to inject and withdraw the
current are either on the top $a$-$b$ face or encompass the ends of the
crystal extending to both opposing $a$-$b$ faces. The features reported here
are common to both geometries. Four more contacts on the same $a$-$b$ face as
the current injection serve to measure the voltage parallel and transverse to
the current flow. The contacts are made by bonding 25~$\mathrm{\mu m}$ gold
wires with silver epoxy fired at $900$~K in an oxygen atmosphere resulting in
contact resistances of less than 3~$\mathrm{\Omega}$ for the current contacts.
The sample is inserted in the bore of a superconducting magnet with the field
along the $c$ direction.

To measure at the high currents required for the free flux flow regime without
significant Joule heating, we apply short (typically 50 $\mu$s or less)
current pulses of isosceles triangular shape with a repetition period of 0.2
to 1 s. Further technical details are given and the issue of Joule heating for
the such experimental conditions is analyzed in Ref.\ \cite{sas2000} with the
conclusion that the temperature change in the area between the voltage
contacts is negligible for the duration of the pulse.

\begin{figure}[ptb]
\includegraphics[width=8.5cm]{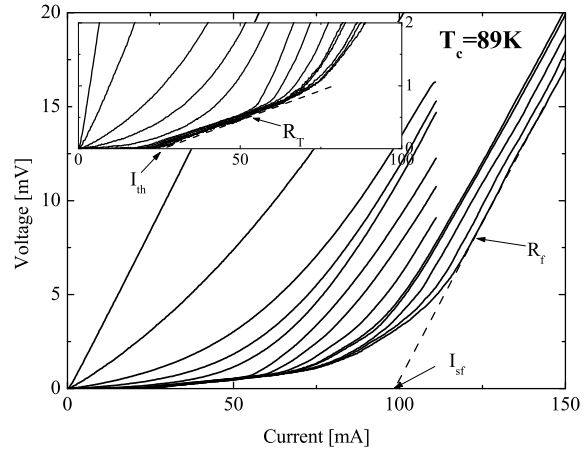} \caption{Typical
voltage-current characteristics in $B = 3$ T at temperatures (from left to
right) 60, 45, 38, 19, 11, 7, and 5 K. Inset: Magnified view of the
low-current region. The dashed lines are linear fits. (See text for the
definition of $I_{sf}$, $I_{th}$, $R_{f}$, and $R_{th}$.)}%
\label{fig:iv}%
\end{figure}

Typical $V$-$I$ characteristics measured at different temperatures and in a
field of 3 T are shown in Fig.\ \ref{fig:iv}. At temperatures below the
vanishing of zero current resistance, which we interpret as the freezing of
the vortex system (in $B=3$~T this happens at $T_{m}=33$~K), dissipation sets
in abruptly beyond a sharply defined threshold current $I_{th}$. Usually, but
not always \cite{sas2000}, a linear segment appears above the threshold
current (see insert) the slope of which we denote by $R_{th}$. At higher
currents, the differential resistance increases further and saturates at a
value $R_{f}\gg R_{th}$ in the high-current limit. The high-current linear
segment of the $V$-$I$ characteristic extrapolates to finite current $I_{f}\gg
I_{th}$ at zero voltage. Above the melting temperature, a finite differential
resistance $R_{I\rightarrow0}$ is observed in the $I\rightarrow0$ limit. The
$V$-$I$ curve, however, is still nonlinear and $I_{sf}$ is finite up to a
temperature $T^{\ast}$ situated between the melting temperature and the
critical temperature $T_{c}$.

\begin{figure}[ptb]
\includegraphics[height=8.5cm, angle=-90]{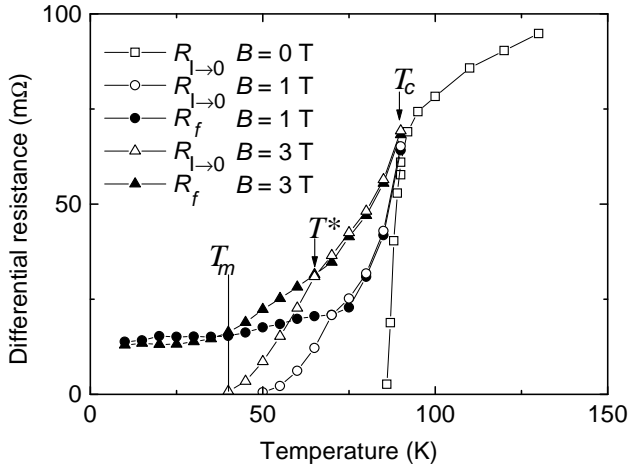} \caption{Temperature
dependence of the low-current ($R_{I \to0}$) and high-current ($R_{f}$)
differential resistances in 1 and 3 T magnetic fields. $R_{I \to0}$ is also
shown for zero field for reference. (See text for the definition of $T_{m}$,
$T_{c}$, and $T^{\ast}$.)}
\label{fig:RvsT}
\end{figure}\begin{figure}[ptbptb]
\includegraphics[height=8.5cm, angle=-90]{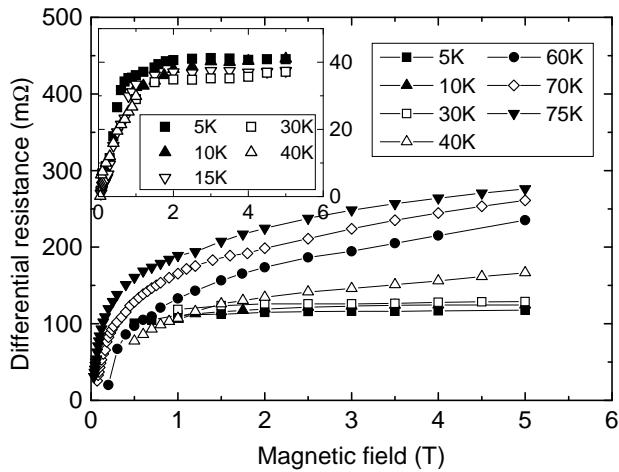} \caption{Magnetic field
dependence of the high-current $R_{f}$ differential resistance at different
temperatures. Insert shows the low-current ($R_{th}$) differential
resistance.}
\label{fig:RvsB}
\end{figure}

The temperature dependence of the low-current and high-current differential
resistances, $R_{I\rightarrow0}$ and $R_{f}$, are shown in
Fig. \ref{fig:RvsT} for two magnetic fields. Between $T_{c}$ and a field
dependent characteristic temperature $T^{\ast}$, these two differential
resistances are identical (the $V$-$I$ curve is linear). At $T^{\ast}$ the two
resistance curves bifurcate; $R_{I\rightarrow0}$ decreases rapidly with
decreasing temperature and reaches zero ($<100$ $\mu \Omega$) at what is taken to be the melting temperature, $T_{m}$, while the
high-current resistance $R_{f}$ remains finite and varies smoothly across the
melting line. The \textquotedblleft pinned liquid\textquotedblright\ domain
extends over $T_{m}<T<T^{\ast}$ where signs of pinning are still present.
Below $T_{m}$, in the vortex solid phase, $R_{I\rightarrow0}$ is zero and
$R_{f}$ attains a finite, temperature-independent value visible in Fig. \ref{fig:RvsT}.
Moreover, this value is the same for 1 T and 3 T and agrees reasonably with a
linear extrapolation of the normal phase resistance to low temperature.

More insight on the low-temperature saturation of the high-current resistance
$R_{f}$ is gained from its field dependence for several temperatures shown in
Fig. \ref{fig:RvsT}. In the liquid phase, we find an approximately $B^{1/2}$ field
dependence over the whole field range. In the solid phase, however, $R_{f}$
becomes independent of magnetic field above about 1 T. On one sample the
measurements were extended to 17 T at 5 K and no variation of $R_{f}$ in
excess of the experimental uncertainties of about 10 \% were found in the
field range 1 to 17 T, in contrast to the 17-fold increase predicted for the
$ab$ resistivity by Eq. (\ref{eq:bs}). The differential resistance close to the
threshold current, $R_{\mathrm{th}}$, behaves similarly (see inset). In fact,
in the parameter range where both quantities can be measured, we find a field
and temperature independent proportionality between $R_{\mathrm{th}}$ and
$R_{f}$.

The behavior of the characteristic currents $I_{\mathrm{th}}$ and
$I_{\mathrm{f}}$ further corroborates the similarities of the dissipation
mechanisms close to threshold and in the high-current limit.  The similarity
is again underlined by the same $B^{-1/2}$ field dependence for both
quantities in the low temperature region.  All these observations suggest that
the same mechanism is responsible for the dissipation in the vicinity of the
threshold current as in the high-current limit. The similarity of behavior is
attributed to the way in which the resistive front propagates from the current
contacts towards the middle of the sample as the current is
increased\cite{currdist}. The current distribution reaches the resistive limit
characterised by $R_{f}$ only after the two resistive fronts meet. As
described earlier\cite{sas2000} at low temperature there is a difference
between field cooled (FC) and zero field cooled (ZFC) preparation. The ZFC
data for both $I_{\mathrm{th}}$ and $I_{\mathrm{f}}$ exhibit a characteristic
peak at the same line in the $(B,T)$ plane which defines a low temperature
region where the FC prepared state is only metastable. Both FC and ZFC
preparations give rise to the same $R_{f}$.

\begin{figure}[ptb]
\includegraphics[width=8.5cm]{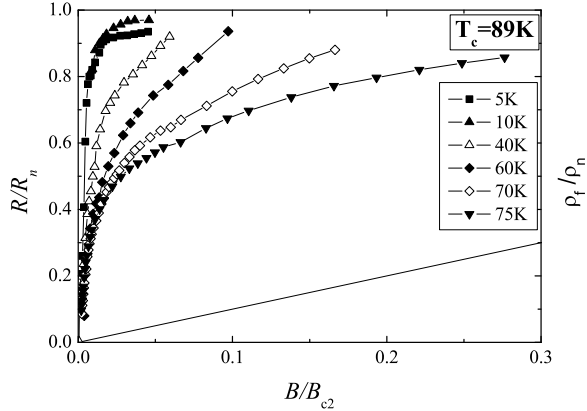} \caption{Reduced
differential resistance ($R_{f}/R_{n}$) as a function of the reduced magnetic
field ($B/B_{c2}$). The BS law for the flux flow resistivity is represented by
a straight line and corresponds to the right hand scale.}
\label{fig:Rperrn}
\end{figure}

A first analysis to get our bearings consists of normalizing the resistance
to the resistance measured in the normal phase and the field to $B_{c2}.$ A
refinement on this is to estimate what the normal resistance would have been
at the temperature of the measurement were the sample not superconducting.
Following this procedure and noting that when the magnetoresistance value
saturates with field its value approximates the expected normal resistance, it
has been hypothesised to be just that and the value used to better interpolate
the normal resistance. For the upper critical field, we use the form
$B_{c2}(T)=(120\ \mathrm{T})[1-(T/T_{c})^{2}]$ yielding $\left.
dB_{c2}/dT\right\vert _{T_{c}}=-2.7$ T/K \cite{li1993}. To estimate $R_{n}$ in
the $T<T_{c}$ range, we use a quadratic interpolating function to connect with
the $T>T_{c}$ range on the hypothesis that the field saturation value for
$R_{f}$ corresponds to $R_{n}(T)$. For the very low field data below about
$0.5$ T where we could not apply sufficient current to reach the $R_{f}$
asymptote, we used the temperature and field independent scaling factor
between $R_{f}$ and $R_{th}$ found at higher fields to estimate $R_{f}$ from
the measurement of $R_{th}.$ The flux flow resistance data treated in this way
are displayed in Fig. \ref{fig:Rperrn}.

It is evident from Fig. \ref{fig:Rperrn} that the ratio of flux flow resistance to normal
resistance is very different from the field proportional BS $\rho_{ab}$
resistivity ratio of Eq. (\ref{eq:bs}), ranging from a $\sqrt{B}$ like variation at low
field to a saturation value for the low temperature vortex solid phase.
Nonetheless there is a trend that with increasing temperature, in the vortex
liquid phase, the saturation feature disappears, at least for the field values
accessible to us. In the solid phase at low temperatures there is strictly no
field dependence of $R_{f}$ whatsoever above about 1 T although as mentioned
above its value is in good agreement with the linear extrapolation to $T=0$ of
the normal resistance measured above $T_{c}$. Since the field where saturation
occurs is in the order of $10^{-2}B_{c2}$, it might appear that for fields
below saturation, the vortices move $\sim10^{2}$ times faster than expected,
or that the vortex friction coefficient is $\sim10^{2}$ smaller, from a simple
application of the BS relation.

\section{Framework for understanding}

The experiment measures resistance and the ratio of resistances is only
proportional to the ratio of resistivities if the current distributions are
identical. This is not the case in anisotropic materials like BSCCO where the
current penetration depth is determined by the anisotropy of the resistivity.
The latter is not only different in the superconducting state but is also
field dependent. From this point of view we might interpret the comparison
made in Fig. \ref{fig:Rperrn} as an indication that the current penetration is considerably
less in the low temperature vortex solid phase, progressively approaching the
normal phase penetration depth at higher temperatures in the vortex liquid
phase. Indeed in the liquid phase, the data are in rough agreement with BS law
if the effects of inhomogeneous current distribution within the sample are
taken into account.

The voltage measured on the top plane of the sample is affected by the
$c$-axis properties because they influence the distribution of the transport
current within the sample \cite{currdist}. In the simplest approach we could
suppose that well into the ohmic response regime the current is distributed as
for a normal anisotropic ohmic conductor\cite{busch}, a surmise which is borne
out by numerical simulations based on the superconductor model of
Ref.\cite{currdist}. The penetration depth $d$ for a length $\ell$ between
current injection points on the surface is then given by $d^{2}\approx
(\sigma^{c}/\sigma^{ab})\ell^{2}$ (provided $d\ll$ sample thickness) and the
resistance $R$ along the length of the sample will scale like $R\approx
\rho^{ab}(\ell/d)w\approx w\sqrt{\rho^{ab}\rho^{c}}$ . If $\rho^{ab}=\rho
_{f}^{ab}$ of Eq. (\ref{eq:bs}) and the resistive response of the Josephson coupled
planes in the superconducting phase is represented by $\rho_{s}^{c}\,$, one
obtains $R_{f}/R_{n}=(\rho_{s}^{c}/\rho_{n}^{c})^{1/2}\sqrt{\alpha B/B_{c2}}$.
Results close to $T_{c}$ are indeed well described by this form. Putting
$\rho_{s}^{c}=\rho_{n}^{c},$ the 75-K data in Fig. \ref{fig:Rperrn}, for instance, are
reasonably well fitted with $\alpha=4$, which, given the uncertainties of the
simple resistive thick-sample estimate, we regard as rough agreement with BS
law. The crux of the matter then, from this viewpoint, is in $\rho_{s}%
^{c}(B,T).$

\begin{figure}
[ptb]
\begin{center}
\includegraphics[width=8.5cm]{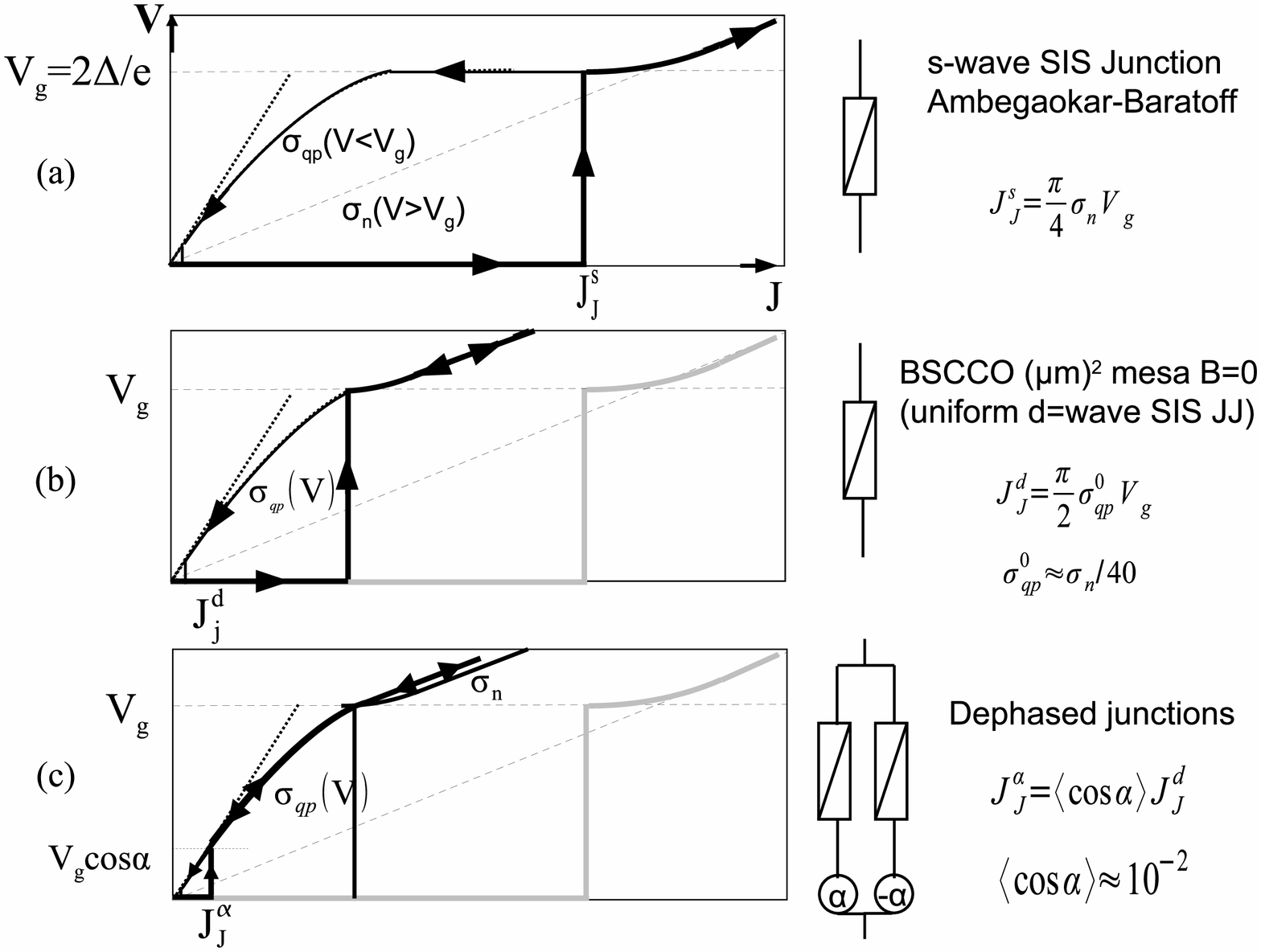}
\caption{Schematic illustration of voltage-current response between Josephson
coupled planes. Part (a) shows a standard SIS junction between s-wave
superconductor planes. (b) shows d-wave response as found in the experiment of
Ref.\cite{Latyshev} where the d-node quasi particle shunt conductance
$\sigma_{qp}^{c}\approx\sigma_{n}^{c}/40$. (c) shows the effect of spatial
dephasing across the junction.}
\label{Josephson junction}
\end{center}
\end{figure}

The $c$ axis conductivity $\sigma_{s}^{c}$ can be obtained from the $I-V$
characteristics of the interplane Josephson junctions. The usual
representation of a Josephson junction in a classical s-wave superconductor
subject to a spatially uniform phase difference is illustrated on Fig. \ref{Josephson junction}a: the voltage response to current is nil up to the Josephson current $J_{J}^{s}$ at
which point the junction opens to $V=V_{g}=2\Delta/e$ and responds to further
current according to the normal phase tunnel conductance $\sigma_{n}%
=1/\rho_{c}^{n}$ which is related to the critical current by the
Ambegaokar-Baratoff relation $J_{J}^{s}=\pi\sigma_{n}\Delta/2e=(\pi
/4)\sigma_{n}V_{g}$ . For currents $J\gg J_{J}^{s}$ the $c$ axis resistivity is
that apppropriate to the normal state $\rho_{s}^{c}=\rho_{n}^{c}.$ Assuming
the penetration depth to be limited by the resistive anisotropy gives the
result quoted above, but this regime would only be attained for extreme
currents $\sim10000$ Acm$^{-2}$ corresponding to $I\geqslant20$ A for our
samples. That is clearly not the regime which concerns us here.

The answer to this problem seems to be that the Josephson critical currents
are much lower than the Ambegoakar-Baratoff prediction. Measurements
\cite{Latyshev} on micrometer size mesa stacks of BSCCO junctions at zero
field show much lower critical currents, $J_{J}^{d}\approx(\pi/2)\sigma
_{qp}V_{g}$ where the quasi-particle conductivity $\sigma_{qp}\approx
\sigma_{n}/40$ as $V\rightarrow0$ at low temperature and $V_{g}\approx50$ mV
yielding $J_{J}^{d}\approx500$ Acm$^{-2}.$ As illustrated schematically on
Fig. \ref{Josephson junction}b the opening of the junction is followed by a voltage response at higher
currents with slope $\rho_{n}$ when $J>J_{J}^{s}\gg J_{J}^{d}.$ But even this
intermediate r\'{e}gime is not observed: a penetration depth of $150$ nm
corresponding to an anisotropy factor $\gamma\sim3000$ would involve $100$
planes and a voltage response of $5$ V, still a factor of $\sim10^{2}$ higher
than observed in our measurements. To understand what we see, we must take
account of the non-uniformity of the phase across the planes due to the random
positioning of vortices from plane to plane. The critical current for the
junction is then reduced to $J_{J}^{\alpha}=\left\langle \cos\alpha
(\mathbf{r})\right\rangle $ $J_{J}^{d}$ where $\alpha(\mathbf{r})$ represents
the phase difference across the junction at position $\mathbf{r}$ in the plane
and $\left\langle \cos\alpha(\mathbf{r})\right\rangle $ the value of
$\cos\alpha$ averaged over the junction area on a Josephson penetration length
scale\cite{shib, gaif}. It is typically of order $10^{-2}.$ As illustrated on
Fig. \ref{Josephson junction}c, the junction will then open to a potential difference of $\left\langle
\cos\alpha(\mathbf{r})\right\rangle V_{g} \ll V_{g}$ due to phase
slippage at $J=J_{J}^{\alpha}$, beyond which it should continue to see a
tunneling conductance $\sigma_{qp}^{c}$ of quasi-particles in the gap until
$V>V_{g},$ when the tunneling of quasi particles at the gap edge will result
in the usual normal phase dynamic conductance. The gap quasi-particle
conductance is thus experienced over a wide current range between
$J_{J}^{\alpha}$ and $J_{J}^{\alpha}/\left\langle \cos\alpha(\mathbf{r}%
)\right\rangle $ over which the resistance ratio is considerably enhanced to
$R_{f}/R_{n}=(\sigma_{n}^{c}/\sigma_{qp}^{c})^{1/2}\sqrt{\alpha B/B_{c2}\text{
}}$ where $(\sigma_{n}^{c}/\sigma_{qp}^{c})\approx30-40$ at low
temperature\cite{Latyshev}, again if we assume Eq. (\ref{eq:bs}) to be valid for the second
square root term. The magnetic field dependence of $\sigma_{qp}\simeq
\sigma_{qp}^{0}(1+\beta(T)B/B_{c2})$ was measured in a subsequent
experiment\cite{Morozov}, also on micrometer size mesa samples. If finally we
include also the effect of quasi-particle conductance $\sigma_{qp}^{ab}(B)$ in
the $ab$ plane, we find
\begin{equation}
\frac{R_{f}}{R_{n}}=\sqrt{\frac{\sigma_{n}^{c}}{\sigma_{qp}^{c}}}\sqrt{\frac{\alpha B/B_{c2}}{(1+\beta B/B_{c2})(1+(\sigma_{qp}^{ab}/\sigma_{n}^{ab})B/B_{c2})}}
\label{eq:RfperRn}
\end{equation}

\begin{figure*}
[ptb]
\begin{center}
\subfigure[]{
\includegraphics[width=8.5cm]{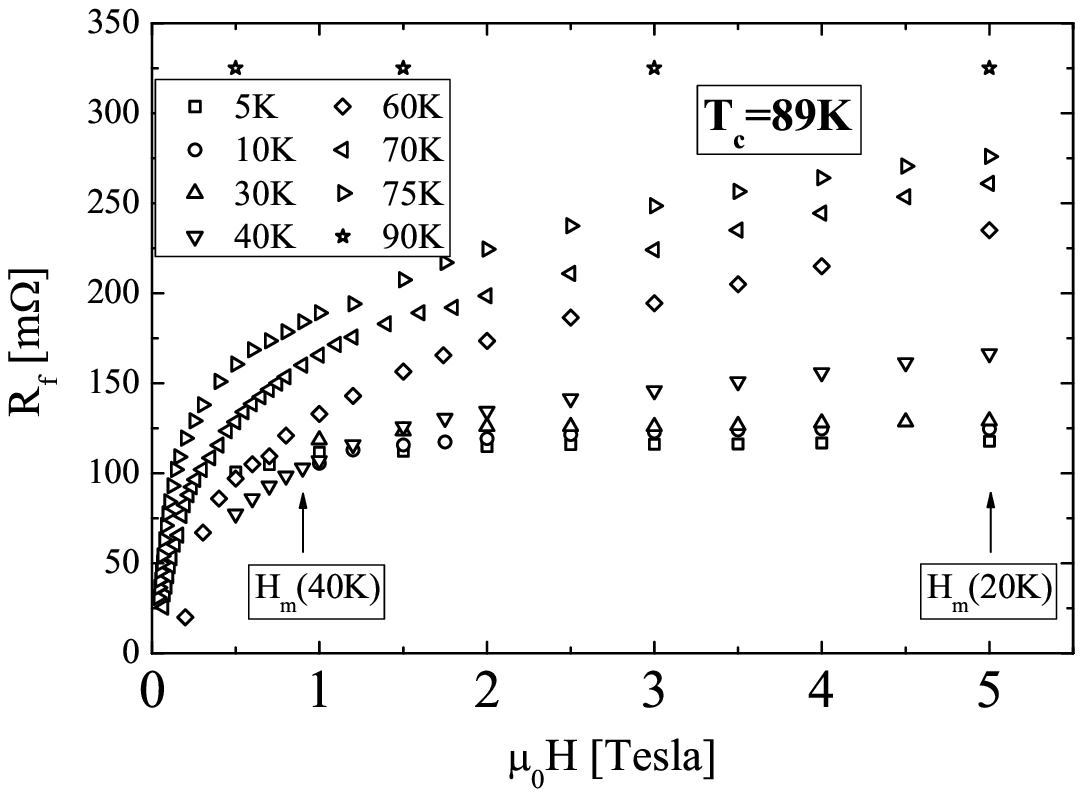}}
\subfigure[]{
\includegraphics[width=8.5cm]{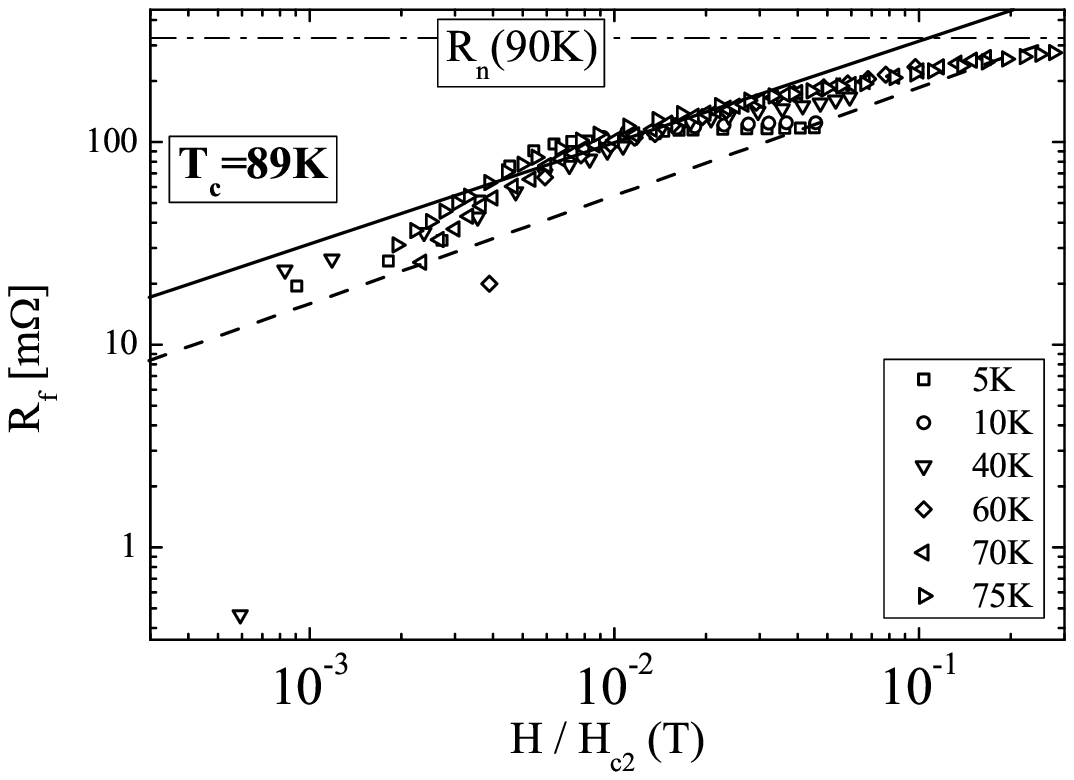}}
\subfigure[]{
\includegraphics[width=8.5cm]{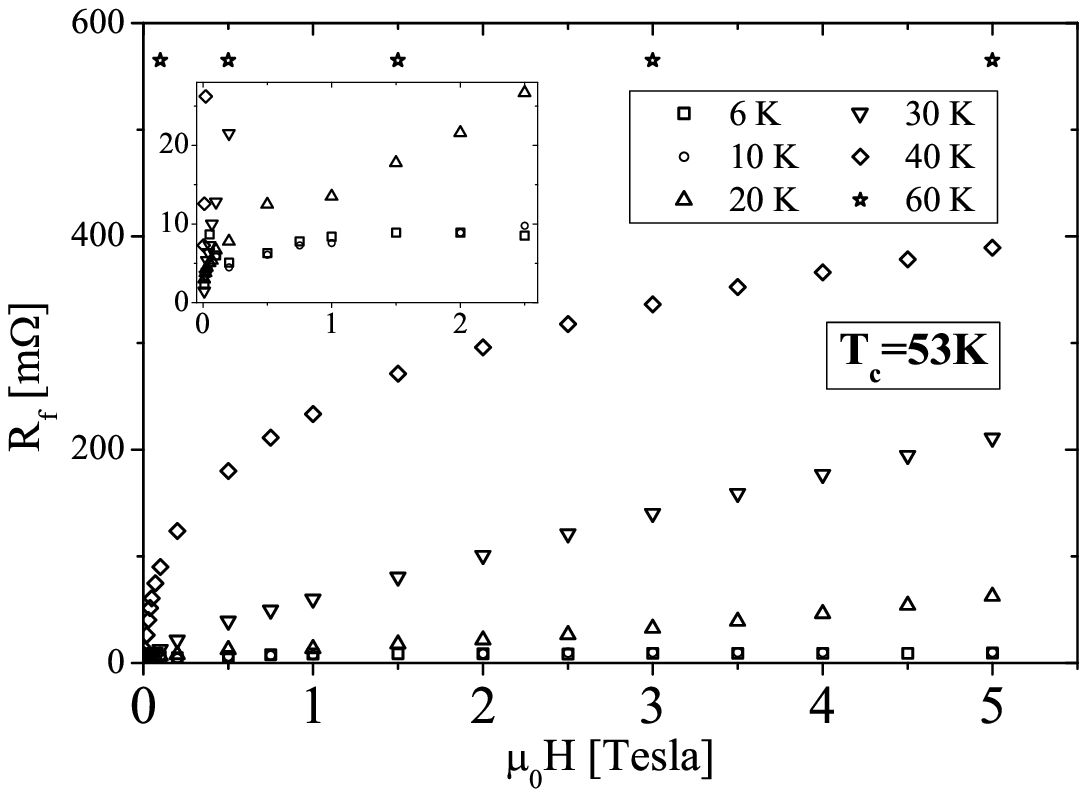}}
\subfigure[]{
\includegraphics[width=8.5cm]{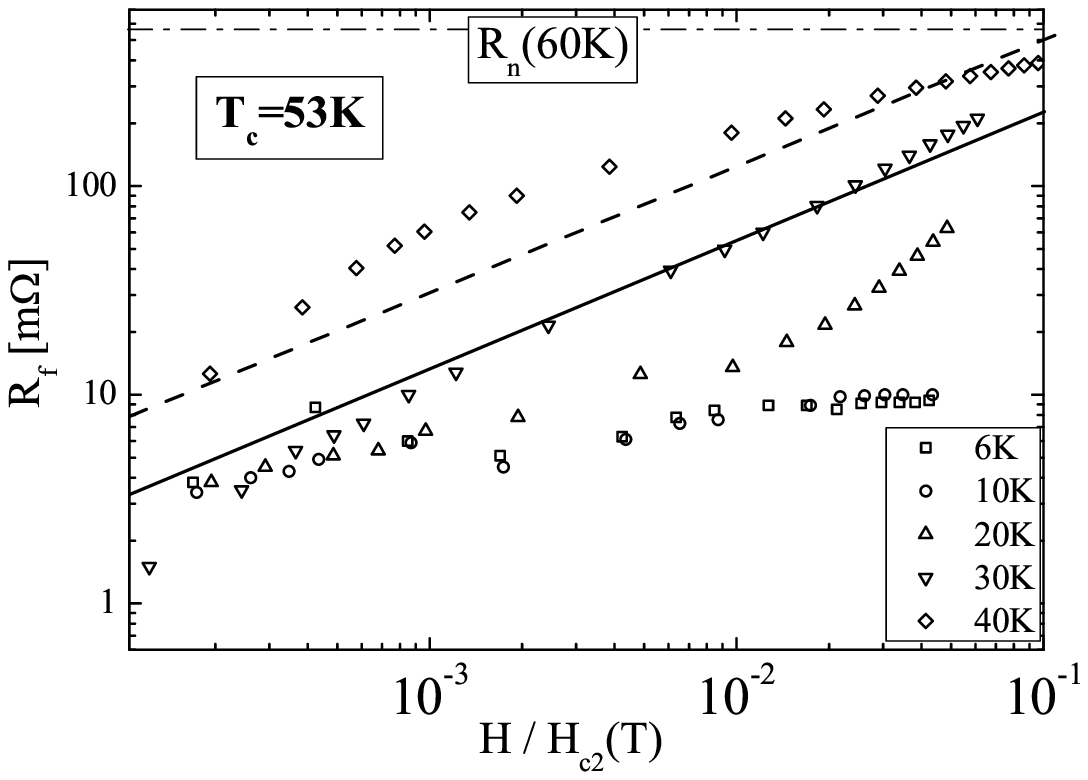}}
\caption{Comparison of experimental measurements with Eq. (\ref{eq:RfperRn}) for the flux flow
resistance as determined by anisotropy using the Bardeen-Stephen flux flow
relation. The raw experimental results are given in a linear representation
(a) and (c) and in a reduced field representation which takes account of the
temperature variation of $B_{c2}=B_{c2}(T)=120(1-(T/T_{c})^{2}$
 T in a log-log representation in (b) and (d). The dashed
lines represent the formula of Eq. (\ref{eq:RfperRn}) using the parameters indicated in the
text. The same parameters are used for both optimal and underdoped samples
for want of data specific to the underdoped case. The solid lines are  simply
an indication of a ($B/B_{c2})^{1/2}$ power dependence.}
\label{Comparison}
\end{center}
\end{figure*}

This is the expression that we compare with the experimental results on Figs. \ref{Comparison}(a)-(d).

\section{Interpretation of results}

Equation (\ref{eq:RfperRn}) is compared with the experimental findings for both the optimal
doped and the underdoped samples on Figs. \ref{Comparison}. For simplicity a single plot of
Eq. (\ref{eq:RfperRn}) is drawn as a dashed line on the log-log plot against
$B/B_{c2}(T)$ where $B_{c2}(T)$ is the temperature corrected value as
described above. It corresponds to a choice of parameter values appropriate to
the optimally doped situation at low temperature $T\rightarrow0$:
$B_{c2}(T=0)=120$~T, $\sigma_{n}^{c}/\sigma_{qp}^{c}=30$\cite{Latyshev},
$\alpha=1$, $\beta=1$\cite{Morozov}, $\sigma_{qp}^{ab}/\sigma_{n}%
^{ab}\rightarrow0$ and $R_{n}/R_{n}(T_{c}^{+})=1/3$ corresponding to a linear
extrapolation to $T=0$ . A parallel solid line is also drawn as the best
$\sqrt{B/B_{c2}}$ fit to the data.

We note that:

1. The order of magnitude is correct to within a factor of 2 for both optimal
and underdoped samples, with the notable exception of the saturation plateau
values in the vortex solid phase, especially in the underdoped sample.

2. The square root behavior predicted by the BS relation is approximately
correct, again with the obvious exception of the plateaus.

3. The magneto-resistance plateaus are all situated in the vortex solid phase.
It is to be remarked that the $20$~K magneto-resistance in the underdoped
sample begins to saturate, but on reaching about $0.5$ T it reverts to an
increasing power law. The melting field for $20$ K is estimated to be about
$0.3$ T from the appearance of finite $R_{I\rightarrow0}$.

Clearly Equation (\ref{eq:RfperRn}) does not describe all the features of the
magneto-resistance, even if it does reasonably well for temperatures above the
vortex solid melting and in the solid region for fields below the saturation.
It has a structure which could in principle describe a plateau if the $\beta$
coefficient in the field dependence of the shunt conductance of the Josephson
coupled layers were to be much larger $(\sim10^{2})$ such as might be
introduced by aligned vortex core NIN tunneling junctions or possibly $d$-wave
node quasi-particle to vortex core tunneling. Such effects have not however
been seen in the mesa magneto-conductivity measurements\cite{Morozov},
although these were performed at relatively high temperature ($>20$ K) and
high field, so in the vortex liquid phase.

\section{Conclusion}

It is difficult to conclude that the BS relation is untrue, either by a
multiplicative factor or in its form. To determine the multiplicative factor
requires a good way of estimating the normal resistivities in the
superconducting phase. Also, because the agreement is reasonably good at high
temperature and the vortex entity is not expected to change nature from high
to low temperature, the serious disagreement at low temperature is much more
likely to be related to the nature of the solid phase as the BS relation is a
single vortex property.

It can however be concluded from the sublinear, approximately $B^{1/2}$, field
dependence that if the linear field dependence of the BS relation is correct,
the current distribution is not constant and its penetration depth is
controlled by the resistive anisotropy and thus by $\rho_{f}$. 

\section{Acknowledgements}

We acknowledge with pleasure fruitful discussions with F.Portier, I.Tutto,
L.Forro and T.Feher and the help and technical expertise of F.Toth. L.Forro
and the EPFL laboratory in Lausanne have contributed in a very essential way
to sample preparation and characterisation. Finally we acknowledge with
gratitude the OTKA funding agency for OTKA grant no. K 62866.

\end{document}